\documentclass{aa}
\usepackage[varg]{txfonts}
 
\begin{document}
 
\title{A thermophysical analysis of the (1862) Apollo Yarkovsky and YORP effects}

\titlerunning{(1862) Apollo Yarkovsky/YORP effects}

\author{B. Rozitis\inst{1}
	\and S. R. Duddy\inst{2}
	\and S. F. Green\inst{1}
	\and S. C. Lowry\inst{2}
} 

\authorrunning{B. Rozitis et al.}

\institute{Planetary and Space Sciences, Department of Physical Sciences, The Open University, Walton Hall, Milton Keynes, MK7 6AA, UK
	\\\email{b.rozitis@open.ac.uk}
	\and Centre for Astrophysics and Planetary Science, School of Physical Sciences (SEPnet), The University of Kent, Canterbury, CT2 7NH, UK
} 

\date{Received XXX / Accepted XXX}

\abstract
{The Yarkovsky effect, which causes orbital drift, and the YORP effect, which causes changes in rotation rate and pole orientation, play important roles in the dynamical and physical evolution of asteroids. Near-Earth asteroid (1862) Apollo has strong detections of both orbital semimajor axis drift and rotational acceleration.} 
{We produce a unified model that can accurately match both observed effects using a single set of thermophysical properties derived from ground-based observations, and we determine Apollo's long term evolution.} 
{We use light-curve shape inversion techniques and the advanced thermophysical model (ATPM) on published light-curve, thermal-infrared, and radar observations to constrain Apollo's thermophysical properties. The derived properties are used to make detailed predictions of Apollo's Yarkovsky and YORP effects, which are then compared with published measurements of orbital drift and rotational acceleration. The ATPM explicitly incorporates 1D heat conduction, shadowing, multiple scattering of sunlight, global self-heating, and rough surface thermal-infrared beaming in the model predictions.} {We find that ATPM can accurately reproduce the light-curve, thermal-infrared, and radar observations of Apollo, and simultaneously match the observed orbital drift and rotational acceleration using: a shape model with axis ratios of 1.94:1.65:1.00, an effective diameter of 1.55 $\pm$ 0.07 km, a geometric albedo of 0.20 $\pm$ 0.02, a thermal inertia of 140 $_{-100}^{+140}$ J m$^{-2}$ K$^{-1}$ s$^{-1/2}$, a highly rough surface, and a bulk density of 2850 $_{-680}^{+480}$ kg m$^{-3}$. Using these properties we predict that Apollo's obliquity is increasing towards the 180\degr~YORP asymptotic state at a rate of 1.5 $_{-0.5}^{+0.3}$ degrees per $10^{5}$ yr.}
{The derived thermal inertia suggests that Apollo has loose regolith material resting on its surface, which is consistent with Apollo undergoing a recent resurfacing event based on its observed Q-type spectrum. The inferred bulk density is consistent with those determined for other S-type asteroids, and suggests that Apollo has a fractured interior. The YORP effect is acting on a much faster timescale than the Yarkovsky effect and will dominate Apollo's long term evolution. The ATPM can readily be applied to other asteroids with similar observational data sets.}

\maketitle

\section{Introduction}
\subsection{The Yarkovsky and YORP effects}

The asymmetric reflection and thermal re-radiation of sunlight from an asteroid’s surface imposes a net force (Yarkovsky effect) and torque (Yarkovsky-O’Keefe-Radzievskii-Paddack or YORP effect) which are of fundamental importance for the dynamical and physical evolution of small asteroids in the solar system [see review by Bottke et al. (2006)].

The Yarkovsky effect results in a drift in the semi-major axis of an asteroid’s orbit. It is invoked as the primary mechanism to deliver asteroids smaller than 40 km in the main belt to resonance zones capable of transporting them to Earth-crossing orbits, and dispersing asteroid families. The Yarkovsky effect has been detected by sensitive radar ranging for (6489) Golevka (Chesley et al. 2003) and (101955) 1999 RQ36 (Chesley et al. 2012), by deviations from predicted ephemerides over a long time span for (152563) 1992 BF (Vokrouhlický et al. 2008) and for 54 other near-Earth asteroids (Chesley et al. 2008; Nugent et al. 2012; Farnocchia et al. 2013), and indirectly through the observed orbital distribution of the Karin cluster asteroid family (Nesvorný \& Bottke 2004). The Yarkovsky effect adds significant uncertainties to predictions of the orbits of potentially hazardous asteroids during very close encounters with the Earth, such as (99942) Apophis (Chesley 2006; Giorgini et al. 2008; Shor et al. 2012), and adds complications for determining the ages of unbound asteroid pairs (Duddy et al. 2012, 2013).

The YORP effect changes an asteroid’s rotation period and the direction of its spin axis. It has been detected through observations of phase shifts in photometric light-curves of four near-Earth Asteroids, (54509) YORP (Lowry et al. 2007; Taylor et al. 2007), (1862) Apollo (Kaasaleinen et al. 2007; Ďurech et al. 2008a), (1620) Geographos (Ďurech et al. 2008b), and (3103) Eger (Ďurech et al. 2012). The YORP effect can explain the excess of very fast and very slow rotators in the population of asteroids smaller than 40 km (Pravec et al. 2008). Continued spin-up of gravitationally bound aggregates (rubble-pile asteroids) will result in changes of shape and/or mass shedding (Holsapple 2010). YORP spin-up is proposed as a viable formation mechanism for binary asteroids as a result of re-aggregation of particles lost from the equator of a fast-spinning asteroid (Walsh et al. 2008). Approximately 15\% of near-Earth asteroids are inferred to be binaries (Pravec et al. 2006). In particular, radar observations of the binary (66391) 1999 KW4 (Ostro et al. 2006) exhibit the typical physical and orbital characteristics predicted by YORP spin-up. It has also been recently suggested that YORP spin-up and fission of contact binary asteroids is a viable formation mechanism of unbound asteroid pairs (Pravec et al. 2010). So far only indirect observational evidence for spin axis change has been found through the clustering of rotation axis directions in asteroid families (Vokrouhlický et al. 2003).

Predictions of the Yarkovsky and YORP effects must take into account the asteroid's size and shape, mass and moment of inertia, surface thermal properties, rotation state, and its orbit about the Sun. Past and current models have neglected or dismissed the effects of global self-heating (i.e. mutual self-heating of interfacing surface elements inside large shape concavities) and rough surface thermal-infrared beaming (i.e. re-radiation of absorbed sunlight back towards the Sun as a result of surface roughness). They have also tended to focus on modelling either the Yarkovsky or YORP effect individually even though they are interdependent. A unified model of both effects is therefore required to accurately predict the long term dynamical evolution of asteroids affected by them. (1862) Apollo is an ideal case to test such a unified model since it is the only asteroid which has strong positive detections of both orbital [mean semi-major axis decrease of 32.1 $\pm$ 3.4 m yr$^{-1}$ from Chesley et al. (2008), Nugent et al. (2012), and Farnocchia et al. (2013)] and spin rate [rotation rate increase of (7.3 $\pm$ 1.6) $\times10^{-3}$ rad yr$^{-2}$ from Ďurech et al. (2008a)] changes as well as light-curve, thermal-infrared, and radar observations to constrain its thermophysical properties. Near-Earth asteroids (54509) YORP and (1620) Geographos also both have observed orbital and spin rate changes but they are less ideal for this kind of analysis. This is because their orbital changes are less well constrained, the shape model of (54509) YORP significantly overestimates the spin rate change, and (1620) Geographos has lower quality thermal-infrared data.

We present results from application of the advanced thermophysical model (ATPM; Rozitis \& Green 2011, 2012, 2013a), which explicitly incorporates 1D heat conduction, shadowing, multiple scattering of sunlight, global self-heating, and rough surface thermal-infrared beaming, to observations of (1862) Apollo (hereafter referred to as just Apollo). ATPM is the first model that can simultaneously interpret thermal-infrared observations and then determine the associated Yarkovsky and YORP effects for the derived thermophysical properties.

\subsection{(1862) Apollo observational history}

Apollo was discovered by Karl Reinmuth in 1932 but it was lost and not recovered until 1973. It made two close approaches with Earth in 1980 and 1982 (0.055 and 0.059 AU respectively) and two more, at slightly greater distances, in 2005 and 2007 (0.075 and 0.071 AU respectively).

During the 1980s close approaches it was extensively observed at optical and thermal-infrared wavelengths, and by radar. Hahn (1983) performed UBVRI and JHK photometric observations, determining colours typical of S-type asteroids and a fast rotation period of 3.065 hours. Harris et al. (1987) obtained photoelectric light-curves which determined Apollo's phase curve relation and that it was also a retrograde rotator. Lebofsky et al. (1981) performed thermal-infrared observations at wavelengths ranging from 4.8 to 20 $\muup$m, which when combined with a standard thermal model (STM; assumes spherical and non-rotating body in instantaneous equilibrium with sunlight) determined an effective spherical diameter ranging from 1.2 to 1.5 km with a geometric albedo of 0.21 $\pm$ 0.02. Radar observations were conducted by the Goldstone (Goldstein et al. 1981) and Arecibo (Ostro et al. 2002) radio telescopes which determined Apollo's average areal diameter to be no more than 1.6 km, and also gave indications of its pole-on silhouette and disk-integrated radar properties.

Between the 1980s and 2000s close approaches, Apollo's orbit was studied in more detail since differences between predictions and observations of its astrometric position were noticed. Using astrometric observations conducted from 1932 to 1982, Ziolkowski (1983) found that by including an unspecified non-gravitational force in the orbital predictions it improved the match with the observations. This non-gravitational force caused Apollo's orbital semi-major axis to shrink but it was uncertain as to what was its cause. It was hypothesised that Apollo might be an extinct comet which is still outgassing or ejecting material. Yeomans (1991) verified this earlier finding by using astrometric observations conducted from 1932 to 1989, and also attempted to explain the orbital perturbation using a cometary outgassing model. Although improved orbital fits to the astrometric data were obtained, a reliable value for the magnitude of this effect could not be determined. During this period Apollo's size and shape was also studied in more detail. De Angelis (1995) determined Apollo to have flattened poles by fitting a triaxial ellipsoid to two sets of light-curves obtained during the 1980s close approaches. Harris (1998) used his near-Earth asteroid thermal model (NEATM) to reanalyse the 1980s thermal-infrared observations and determined an effective spherical diameter of 1.45 km with a geometric albedo of 0.26 and a beaming parameter of 1.15. The derived beaming parameter suggested a non-zero but small surface thermal inertia that is slightly larger than that of (433) Eros, i.e. $\sim$100 to 200 J m$^{-2}$ K$^{-1}$ s$^{-1/2}$ (Müller 2007). Finally, during this period, Apollo was identified as a Q-type asteroid in the Bus and Binzel feature-based spectroscopic taxonomy system (Bus \& Binzel 2002).

During the 2000s close approaches it was again extensively observed at optical and thermal-infrared wavelengths, and by radar. The Arecibo radio telescope obtained several delay-Doppler radar images (www.naic.edu/$\sim$pradar/sched.shtml), and the \textit{Spitzer} space telescope obtained thermal-infrared photometric and spectroscopic measurements (Van Cleve et al. 2006; Mainzer et al. 2007). Unfortunately the results from these investigations have not been published yet. However, it was announced that the radar observations revealed a small moon orbiting Apollo whose diameter is less than one tenth of Apollo's diameter (Ostro et al. 2005). Kaasalainen et al. (2007) detected an acceleration in Apollo's rotation rate that amounted to one extra rotation cycle in just 40 years by observing small rotational phase shifts in light-curves obtained from 1980 to 2005. This was attributed to the asteroid YORP effect since the shape and rotation pole orientation derived by light-curve inversion predicted a YORP rotational acceleration that was consistent with that observed. Ďurech et al. (2008a) refined the YORP rotational acceleration measurement along with Apollo's shape and rotation pole orientation by using additional light-curve observations obtained during the 2007 close approach. Using a simple Yarkovsky analytical model and surface properties consistent with Apollo's beaming parameter, Vokrouhlický et al. (2005) made predictions of the Yarkovsky effect acting on Apollo (as well other near-Earth asteroids) and indicated that it should be easily detected by radar observations conducted during the 2007 close approach. However, they did not consider that it might have already been detected during the 1980s and 1990s. Finally, Chesley et al. (2008) performed a comprehensive search for evidence of the Yarkovsky effect in orbital fits of near-Earth asteroids. Their direct approach involved modelling the Yarkovsky effect as a purely transverse acceleration that is inversely proportional to the square of the heliocentric distance. This permitted a rapid scan of the asteroid catalogue for objects whose orbital fits were markedly improved by the incorporation of the transverse acceleration. Unsurprisingly, Apollo was the third most likely candidate to exhibit an orbital drift with the two more likely asteroids being the ones that had already been Yarkovsky detected at that time. By using astrometric data ranging from 1930 to 2008 the orbital semi-major axis was again found to shrink, which is consistent with the Yarkovsky effect caused by Apollo's retrograde rotation. This finding has been confirmed by Nugent et al. (2012) and by Farnocchia et al. (2013) who both use similar orbital fitting techniques but with added data weighting treatment.

No further observations of Apollo are planned for the near future because it doesn't make another close approach with the Earth until 2021. However, this encounter will be much further away than previous ones since it has a closest approach distance of 0.212 AU.

\section{Thermophysical modelling}
\subsection{Thermal-infrared flux fitting}

To determine Apollo's thermophysical properties we combine the ATPM with the light-curve-derived shape model and spin state (Ďurech et al. 2008a), and compare the model outputs for various thermophysical properties with the thermal-infrared observations obtained from three nights in 1980 (Lebofsky et al. 1981) via chi-squared fitting. The methodology used here is similar to that presented in Wolters et al. (2011) and Lowry et al. (2012). As described in Rozitis \& Green (2011), the ATPM model can be applied to any atmosphereless planetary body whose global shape is represented by the triangular facet formalism. To determine the thermal emission, the model computes the surface temperature variation for each shape model facet during a rotation by solving the 1D heat conduction equation with a surface boundary condition that includes direct and multiple scattered solar radiation, shadowing, and re-absorbed thermal radiation from interfacing facets. The model explicitly includes thermal-infrared beaming from each shape facet by including roughness facets that represent unresolved surface roughness at spatial scales as small as $\sim$1 cm in the global shape model. Any rough surface type can be defined but hemispherical craters are used since they accurately reproduce the directionality of thermal emission from the Moon and are easy to parameterise [see Rozitis \& Green (2011) for more details]. The degree of roughness is characterised by the fraction of surface, $f_{\mathrm R}$, covered by the hemispherical craters, i.e. 100\% coverage implies a maximum RMS slope of 50\degr. The lunar thermal-infrared data are best fit by a $\sim$41\% coverage consistent with a $\sim$32\degr~RMS slope at 1-cm scales. A Planck function is applied to the derived temperatures and summed across visible shape and roughness facets to give the emitted thermal flux as a function of observation wavelength, rotation phase, and various thermophysical properties. The thermal flux contribution from Apollo's small moon is neglected, as it contributes less than 1\% of the total observed flux.

For Apollo, the free parameters to be constrained by fits to the thermal-infrared observations are the effective diameter (i.e. the equivalent diameter of a sphere with the same volume as the irregularly shaped asteroid), geometric albedo, thermal inertia, and surface roughness. The effective diameter, $D$, and geometric albedo, $p_{\mathrm v}$, are related to one another and can be considered a single free parameter. They are related by
\begin{equation}
D=\frac{10^{-H_{\mathrm v}/5}1329}{\sqrt{p_{\mathrm v}}}\text{ km}
\label{(1)}
\end{equation}
where $H_{\mathrm v}$ is the absolute visual magnitude of the asteroid (Fowler \& Chillemi 1992). The geometric albedo is also related to the effective Bond albedo of the asteroid surface, $A_{\text{B\_EFF}}$, by
\begin{equation}
A_{\text{B\_EFF}}=(0.290+0.684G)p_{\mathrm v}
\label{(2)}
\end{equation}
where $G$ is the asteroid phase parameter (Bowell et al. 1989). In this work, we utilise values of 16.384 and 0.24 for the absolute visual magnitude and phase parameter respectively, which are obtained from Pravec et al. (2012). However, except for the geometric albedo, the derived parameters from the ATPM chi-square fitting are relatively insensitive to small differences in these values. Since the degree of surface roughness influences the effective Bond albedo it is related to the Bond albedo of a smooth flat surface, $A_{\mathrm B}$, by
\begin{equation}
A_{\text{B\_EFF}}=f_{\mathrm R}\frac{A_{\mathrm B}}{2-A_{\mathrm B}}+(1-f_{\mathrm R})A_{\mathrm B}
\label{(3)}
\end{equation}
where $f_{\mathrm R}$ is the roughness fraction as before. Therefore, each effective diameter and roughness fraction combination leads to a unique Bond albedo value to be used in the ATPM. However, it is computationally expensive to run ATPM for each value separately. As in Wolters et al. (2011) and Lowry et al. (2012), we run ATPM for just one Bond albedo value and perform a pseudo-correction to the observed flux for different values. This flux correction factor, $FCF$, is given by
\begin{equation}
FCF=\frac{1-A_{\mathrm B}}{1-A_{\text{B\_MOD}}}
\label{(4)}
\end{equation}
where $A_{\mathrm B}$ is the smooth surface Bond albedo determined by inversion of equation (3) and $A_{\text{B\_MOD}}$ is the model Bond albedo used in the ATPM. A value of $A_{\text{B\_MOD}}=0.1$ is assumed, which results in flux correction factors that are within 10\% of unity. A thermal-infrared emissivity of 0.9 is assumed, and separate thermophysical models were run for thermal inertia values, $\Gamma$, ranging from 0 to 1000 J m$^{-2}$ K$^{-1}$ s$^{-1/2}$ in equally spaced steps using the light-curve-derived shape model and spin state. Similarly, the effective diameter and roughness fraction are also stepped through their plausible ranges, which form a 3-dimensional grid of model test parameters (or test clones) with the thermal inertia steps. Using 650 time steps and 56 depth steps to solve the 1D heat conduction equation, the ATPM typically takes between 10 and 100 revolutions to reach convergence.

The model thermal flux predictions, $F_{\text{MOD}}(\lambda_{n}, \varphi_{n}, \Gamma, D,f_{\mathrm R})$, were then compared with the observations, $F_{\text{OBS}}(\lambda_{n},\varphi_{n})$, and observational errors, $\sigma_{\text{OBS}}(\lambda_{n},\varphi_{n})$, by varying the effective diameter, thermal inertia, and roughness fraction to give the minimum-chi-squared fit
\begin{equation}
\chi^{2}=\sum_{n=1}^{N}\Bigg [\frac{FCF(D,f_{\mathrm R})F_{\text{MOD}}(\lambda_{n},\varphi_{n},\Gamma,D,f_{\mathrm R})-F_{\text{OBS}}(\lambda_{n},\varphi_{n})}{\sigma_{\text{OBS}}(\lambda_{n},\varphi_{n})}\Bigg]^{2}
\label{(5)}
\end{equation}
for a set of $N$ observations with wavelength $\lambda_{n}$ and rotation phase $\varphi_{n}$. Table 1 summarises the three sets of thermal-infrared observations, and Table 2 summarises the fixed model parameters, used to determine Apollo's thermophysical properties. The model fitting required knowledge of the exact rotation phase of Apollo at which each observation was made. Due to Apollo's fast rotation, the observation date given to 2 decimal places by Lebofsky et al. (1981) unfortunately does not give sufficient time resolution. This corresponds to an uncertainty of $\pm$14\degr~in rotation phase for each flux measurement, and the fitting procedure takes this into account by selecting the rotational phase that best matches each individual flux measurement within the allowed range. Due to the relatively low number of thermal flux measurements and their rotational phase uncertainties, we chose a parameter region bounded by a constant $\Delta\chi^{2}$ value at the 3-$\sigma$ confidence level to define the range of possible parameters inside the 3-dimensional test grid. All three sets of thermal-infrared observations are fitted simultaneously in the ATPM chi-square fitting.

\subsection{Results and shape optimisation}

Using the `original' Ďurech et al. (2008a) light-curve-derived shape model, the ATPM chi-square fitting determines an effective diameter of 1.68 $\pm$ 0.09 km (i.e. mean value and standard deviation), a geometric albedo of 0.17 $\pm$ 0.02, a thermal inertia of 200 $_{-140}^{+220}$ J m$^{-2}$ K$^{-1}$ s$^{-1/2}$ (i.e. median value and 1-$\sigma$ spread), and a roughness fraction of 60 $\pm$ 30 \% for Apollo. This effective diameter is slightly larger than 1.45 km determined by NEATM (Harris 1998), which has a nominal uncertainty of $\sim$15\% due to simple-model assumptions. Using the relative shape model dimensions, this effective diameter implies a maximum equatorial diameter, $A$, of 2.43 $\pm$ 0.13 km, which can be compared against that determined by Doppler radar measurements. Combining Apollo's observed Doppler radar bandwidth (Ostro et al. 2002) with the light-curve-derived pole orientation, which has an uncertainty of $\pm$7\degr~in arc (Kaasalainen et al. 2007; Ďurech et al. 2008a), gives Apollo's maximum equatorial diameter as 2.07 $\pm$ 0.07 km (see Table 1 for the radar observation geometry). The ATPM fit therefore overestimates the maximum equatorial diameter by a factor of 1.17 $\pm$ 0.07.

According to Ďurech et al. (2008a), the z-axis of the Apollo shape model was not well constrained by the light-curve photometry, allowing some ambiguity in the real shape. Since the sub-Earth latitude of the thermal-infrared observations was low, a simple stretch in the shape model z-axis by a factor of 1.17 could reconcile the thermal-infrared and radar derived diameters. Indeed, ATPM chi-square fitting with this `stretched' shape model determines an effective diameter of 1.54 $\pm$ 0.07 km and a maximum equatorial diameter of 2.12 $\pm$ 0.09 km, which are in much better agreement with the NEATM and radar derived values respectively. A tighter constraint on the thermal inertia value is also obtained with 140 $_{-100}^{+160}$ J m$^{-2}$ K$^{-1}$ s$^{-1/2}$, and the geometric albedo and roughness fraction are determined to be 0.21 $\pm$ 0.02 and 60 $\pm$ 30 \% respectively. However, the `stretched' shape model produces a fit to the light-curve observations that results in a relative-chi-square value [see equation 7 of Kaasalainen \& Torppa (2001)] $\sim$30\% higher than that produced by the `original' shape model. Usually, a tolerance within 10\% of the best relative-chi-square value is allowed (e.g. Ďurech et al. 2008a), and so the `stretched' shape model is deemed not to be ideal given this criteria.

To find an idealised shape model that optimises the fit to all of the light-curve, thermal-infrared, and radar observations, we produced a series of Apollo shape model variants with different relative shape dimensions using a YORP-modified version of the Kaasalainen light-curve inversion software (Kaasalainen \& Torppa 2001; Kaasalainen et al. 2001), as the original release did not include YORP capability. In particular, we sought a shape model with a maximum equatorial diameter to effective diameter ratio, $A/D$, of 1.37, which is equal to that of the `stretched' shape model (the `original' shape model has $A/D$ of 1.45). We found that the software's convexity regularisation parameter and iteration number dictated the $A/D$ value of the shape model the software would produce. The convexity regularisation parameter is used to address the issues of albedo variation across the asteroid surface and shape concavities in the light-curve shape inversion. The software does this by introducing a small dark facet to ensure that the sum of products of the facet areas and facet normals exactly equals zero, and its area is proportional to the convexity regularisation parameter. The iteration number is the set number of times the software's algorithm is run in order to find the optimal shape solution. As shown in Fig. 1, the $A/D$ value converges with iteration number but the value at which it converges at is dictated by the convexity regularisation parameter. We chose a convexity regularisation parameter of 0.8 and iteration number of 475 since they produce a shape model that converges on the required $A/D$ value. All other shape modelling parameters were kept the same as that used by Ďurech et al. (2008a).

Repeating the ATPM chi-square fitting with the `new' shape model determines an effective diameter of 1.55 $\pm$ 0.07 km, a maximum equatorial diameter of 2.13 $\pm$ 0.09 km, a geometric albedo of 0.20 $\pm$ 0.02, a thermal inertia of 140 $_{-100}^{+140}$ J m$^{-2}$ K$^{-1}$ s$^{-1/2}$, and a roughness fraction of 60 $\pm$ 30 \%. These results are very similar to those obtained by the `stretched' shape model, and the light-curve relative-chi-square fit is within 10\% of that of the `original' shape model. Fig. 2 provides a visual comparison of the `original', `stretched', and `new' shape models; Table 3 summarises their ATPM derived thermophysical properties; and Fig. 3 shows an example ATPM fit to the thermal-infrared observations using the `new' shape model.

Fig. 4 shows the distribution of possible thermal inertia values derived by ATPM chi-square fitting at the 3-$\sigma$ confidence level for the three different shape model variants; and the co-variance of the average fitted effective diameter, geometric albedo, and roughness fraction with thermal inertia. The thermal inertia distribution is obtained by counting each allowed test clone with a specific thermal inertia value and dividing by the total number of allowed test clones ($\sim$10$^{3}$ to 10$^{4}$ clones). The co-variance of the other parameters with thermal inertia is obtained by averaging the values of the allowed test clones in each thermal inertia bin. Later distributions and parameter co-variances presented in this work are calculated in a similar way. For the `new' shape model variant, the most likely thermal inertia value is at $\sim$80 J m$^{-2}$ K$^{-1}$ s$^{-1/2}$, but thermal inertia values up to $\sim$580 J m$^{-2}$ K$^{-1}$ s$^{-1/2}$ cannot be ruled out. For comparison purposes, the thermal inertia value of lunar regolith is $\sim$50 J m$^{-2}$ K$^{-1}$ s$^{-1/2}$ [see Rozitis \& Green (2011) and references therein], the average value determined for km-sized near-Earth asteroids is 200 $\pm$ 40 J m$^{-2}$ K$^{-1}$ s$^{-1/2}$ (Delbo et al. 2007), and the average value determined for binary near-Earth asteroids is 480 $\pm$ 70 J m$^{-2}$ K$^{-1}$ s$^{-1/2}$ (Delbo et al. 2011). This level of thermal inertia suggests that loose regolith material exists on the surface of Apollo, and has a larger grain size than that of lunar regolith. The presence of loose regolith material is also suggested by the fact that Apollo is a Q-type asteroid, as such asteroids are believed to acquire their `unweathered' spectra through a resurfacing event caused by a recent close planetary encounter (Binzel et al. 2010). In particular, Nesvorný et al. (2010) show that within the past 0.5 Myr Apollo's median closest encounter with Venus was only $\sim$9 planetary radii making it a very good candidate for having a recent resurfacing event. The average fitted effective diameter and roughness fraction both increase with increasing thermal inertia, and the geometric albedo decreases with increasing thermal inertia.

\section{Yarkovsky and YORP modelling}

Using the three different shape model variants, we predict the Yarkovsky and YORP effects acting on Apollo for the range of possible thermophysical properties (or allowed test clones) determined by the thermal-infrared flux fitting. The Yarkovsky and YORP effects can be determined by computing the total recoil forces and torques from reflected and thermally emitted photons from the asteroid surface. These calculations are made for both the shape and roughness facets, and are averaged over both the asteroid rotation and elliptical orbit [see Rozitis \& Green (2012, 2013a) for methodology]. As previously shown in Rozitis \& Green (2012), the inclusion of rough surface thermal-infrared beaming in the predictions, on average, enhances the Yarkovsky orbital drift whilst it dampens the YORP rotational acceleration by orders of several tens of per cent. The model predictions are then compared against published Yarkovsky semi-major axis drift and YORP rotational acceleration measurements. To ensure that these predictions are as accurate as possible we investigated how small differences in solving the 1D heat conduction equation affected the model results, and we briefly summarise the important findings in Appendix A.

\subsection{Yarkovsky effect}

Apollo has three measurements of Yarkovsky semi-major axis drift: -36.5 $\pm$ 3.9 m yr$^{-1}$ by Chesley et al. (2008), -34.4 $\pm$ 3.0 m yr$^{-1}$ by Nugent et al. (2012), and -25.4 $\pm$ 3.9 m yr$^{-1}$ by Farnocchia et al. (2013). We combine the three measurements to produce an average drift of -32.1 m yr$^{-1}$ with a standard error of 3.4 m yr$^{-1}$, which we use to produce a normal distribution of possible drifts that has a mean and standard deviation equal to these values. For model comparison, 200 values of possible Yarkovsky drift are randomly selected from this distribution that range from -21.8 to -43.1 m yr$^{-1}$, which ensures that the three measured values are encompassed. 

Since the best fit effective diameter, geometric albedo, and roughness fraction varies with thermal inertia, their co-variance must be taken into account within the Yarkovsky drift predictions. To evaluate the diurnal Yarkovsky drift, the model is run twice for each thermal inertia value: once with a completely smooth surface, $\mathrm d a/\mathrm d t(\Gamma)_{\text{smooth}}$, and once with a completely rough surface, $\mathrm d a/\mathrm d t(\Gamma)_{\text{rough}}$, with fixed effective diameter and bulk density, $D_{0}$ and $\rho_{0}$. The seasonal Yarkovsky drift, $\mathrm d a/\mathrm d t(\Gamma)_{\text{seasonal}}$, is evaluated separately and is assumed to be equally applicable to both a smooth and rough surface since the seasonal thermal skin depth ($\sim$1 to 10 m) is much greater than the typical surface roughness spatial scale ($\sim$1 cm). The overall Yarkovsky drift for other properties, $\mathrm d a/\mathrm d t(\Gamma,D,f_{\mathrm R},\rho)$, can then be calculated using
\begin{equation}
\begin{split}
\frac{\mathrm d a}{\mathrm d t}(\Gamma,D,f_{\mathrm R},\rho)=&\Bigg(\frac{D_{0}}{D}\Bigg)\Bigg(\frac{\rho_{0}}{\rho}\Bigg)FCF(D,f_{\mathrm R})\Bigg[(1-f_{\mathrm R})\frac{\mathrm d a}{\mathrm d t}(\Gamma)_{\text{smooth}}\\
&+f_{\mathrm R}\frac{\mathrm d a}{\mathrm d t}(\Gamma)_{\text{rough}}+\frac{\mathrm d a}{\mathrm d t}(\Gamma)_{\text{seasonal}}\Bigg]
\end{split}
\label{(6)}
\end{equation}
where the flux correction factor takes into account changes in absorbed/emitted flux caused by different Bond albedo values. Furthermore, the bulk density for a set of properties, $\rho(\Gamma,D,f_{\mathrm R})$ can be determined from the measured semi-major axis drift, $\mathrm d a/\mathrm d t_{\text{measured}}$, using
\begin{equation}
\rho(\Gamma,D,f_{\mathrm R})=\rho_{0}\Bigg(\frac{\mathrm d a}{\mathrm d t}(\Gamma,D,f_{\mathrm R},\rho_{0})\Bigg/\frac{\mathrm d a}{\mathrm d t}_{\text{measured}}\Bigg)\text{ .}
\label{(7)}
\end{equation}
The Yarkovsky effect acting on each Apollo shape variant can be assessed by using these relations and the ranges of allowed parameters (or test clones) that were derived at the 3-$\sigma$ confidence level from the thermal-infrared flux fitting. Fig. 5a shows the average Yarkovsky drift as a function of thermal inertia with fixed bulk density. As indicated, the Yarkovsky drift is strongest at a thermal inertia value of $\sim$200 J m$^{-2}$ K$^{-1}$ s$^{-1/2}$ for all three shape variants. Fig. 5b shows the average bulk density required to match the observed orbital drift as a function of thermal inertia, which ranges from $\sim$1000 up to $\sim$3100 kg m$^{-3}$. Lastly, Fig. 5c shows the distribution of possible bulk densities derived for each of the three different shape variants. The derived bulk densities have median values and 1-$\sigma$ spreads of 2400 $_{-470}^{+450}$, 2820 $_{-660}^{+480}$, and 2850 $_{-680}^{+480}$ kg m$^{-3}$ for the `original', `stretched', and `new' shape variants respectively. The `stretched' and `new' shape model bulk densities are very similar to one another, and are also very similar to the average value of 2720 $\pm$ 540 kg m$^{-3}$ determined for S-type asteroids (Carry 2012). Table 4 summarises Apollo's density and mass properties for the three different shape variants investigated.

\subsection{YORP effect}

Apollo has two measurements of YORP rotational acceleration: (7.1 $\pm$ 1.7) $\times10^{-3}$ rad yr$^{-2}$ by Kaasalainen et al. (2007), and (7.3 $\pm$ 1.6) $\times10^{-3}$ rad yr$^{-2}$ by Ďurech et al. (2008a). We use the value determined by Ďurech et al. (2008a) for model comparison since it is an update of the value determined by Kaasalainen et al. (2007).

When the diurnal Yarkovsky drift is evaluated for each thermal inertia value, as described above, the ATPM simultaneously evaluates the YORP effect acting on the same asteroid. There are two YORP components of interest: the rotational acceleration, $\mathrm d \omega/\mathrm d t$, and the rate of change in obliquity, $\mathrm d \xi/\mathrm d t$. The overall YORP rotational acceleration as a function of the asteroid properties, $\mathrm d \omega/\mathrm d t(D,f_{\mathrm R},\rho)$, can be calculated using
\begin{equation}
\frac{\mathrm d \omega}{\mathrm d t}(D,f_{\mathrm R},\rho)=\Bigg(\frac{D_{0}}{D}\Bigg)^{2}\Bigg(\frac{\rho_{0}}{\rho}\Bigg)\Bigg[(1-f_{\mathrm R})\frac{\mathrm d \omega}{\mathrm d t}_{\text{smooth}}+f_{\mathrm R}\frac{\mathrm d \omega}{\mathrm d t}_{\text{rough}}\Bigg]
\label{(8)}
\end{equation}
where $\mathrm d \omega/\mathrm d t_{\text{smooth}}$ and $\mathrm d \omega/\mathrm d t_{\text{rough}}$ are the YORP rotational acceleration values for a smooth and rough surface, respectively, which are independent of albedo (and hence flux correction factor) and thermal inertia. However, the thermally emitted component of the YORP obliquity shift, $\mathrm d \xi_{\text{RAD}}/\mathrm d t(\Gamma,D,f_{\mathrm R},\rho)$, is dependent on albedo and thermal inertia, and can be calculated using
\begin{equation}
\begin{split}
\frac{\mathrm d \xi_{\text{RAD}}}{\mathrm d t}(\Gamma,D,f_{\mathrm R},\rho)=&\Bigg(\frac{D_{0}}{D}\Bigg)^{2}\Bigg(\frac{\rho_{0}}{\rho}\Bigg)\Bigg[(1-f_{\mathrm R})\frac{\mathrm d \xi_{\text{RAD}}}{\mathrm d t}(\Gamma)_{\text{smooth}}\\
&+f_{\mathrm R}\frac{\mathrm d \xi_{\text{RAD}}}{\mathrm d t}(\Gamma)_{\text{rough}}\Bigg]
\end{split}
\label{(9)}
\end{equation}
where $\mathrm d \xi_{\text{RAD}}/\mathrm d t(\Gamma)_{\text{smooth}}$ and $\mathrm d \xi_{\text{RAD}}/\mathrm d t(\Gamma)_{\text{rough}}$ are the rates of YORP obliquity shift for a smooth and rough surface respectively. The overall rate of YORP obliquity shift can be found by combining the thermal component with the scattered component, $\mathrm d \xi_{\text{SCAT}}/\mathrm d t(D,f_{\mathrm R},\rho)$, using
\begin{equation}
\begin{split}
\frac{\mathrm d \xi}{\mathrm d t}(\Gamma,D,f_{\mathrm R},\rho)=&FCF(D,f_{\mathrm R})\frac{\mathrm d \xi_{\text{RAD}}}{\mathrm d t}(\Gamma,D,f_{\mathrm R},\rho)\\
&+SCF(D,f_{\mathrm R})\frac{\mathrm d \xi_{\text{SCAT}}}{\mathrm d t}(D,f_{\mathrm R},\rho)
\end{split}
\label{(10)}
\end{equation}
where $FCF$ is the flux correction factor as before for adjusting the amount of absorbed/emitted flux, and $SCF$ is another flux correction factor to adjust the amount of scattered flux. It is related to the flux correction factor and the model Bond albedo, $A_{\text{B\_MOD}}$, via
\begin{equation}
SCF=\frac{1+A_{\text{B\_MOD}}FCF-FCF}{A_{\text{B\_MOD}}}\text{ .}
\label{(11)}
\end{equation}

Using these relations, and the 3-$\sigma$ ranges of allowed parameters (or test clones) determined from the thermal-infrared flux fitting, and the bulk densities determined from the Yarkovsky effect modelling, allows the YORP effect acting on each Apollo shape variant to be assessed. Fig. 6a shows the average YORP rotational acceleration as a function of thermal inertia. As previously shown in other works (e.g. Čapek \& Vokrouhlický 2004), the YORP rotational acceleration is independent of thermal inertia when other properties that influence its strength are kept fixed. However, as shown here, the average YORP rotational acceleration does vary with thermal inertia because the best fit effective diameter and roughness fraction determined from the thermal-infrared flux fitting vary with thermal inertia (see Fig. 4). Furthermore, the most likely thermal inertia values (see Fig. 4) produce YORP rotational acceleration values that fall in the middle of the measured range for each of the three different shape variants. Fig. 6b shows the distribution of possible YORP rotational accelerations derived for each shape variant, which all peak at the lower end of the measured range. The derived YORP rotational accelerations have median values and 1-$\sigma$ spreads of (6.8 $_{-1.2}^{+2.7}$), (6.8 $_{-1.3}^{+3.5}$), and (6.1 $_{-1.2}^{+3.2}$) $\times10^{-3}$ rad yr$^{-2}$ for the `original', `stretched', and `new' shape variants respectively, which all agree well with the measured value of (7.3 $\pm$ 1.6) $\times10^{-3}$ rad yr$^{-2}$. Fig. 6c shows the average rate of YORP obliquity shift as a function of thermal inertia, and Fig. 6d shows the distribution of possible values. As shown, the obliquity of Apollo is increasing for each shape variant, which indicates that it is approaching one of the YORP asymptotic states at 180\degr~obliquity (Čapek \& Vokrouhlický 2004). The derived rates of YORP obliquity shift have median and 1-$\sigma$ spreads of 2.4 $_{-0.6}^{+0.5}$, 1.7 $_{-0.5}^{+0.3}$, and 1.5 $_{-0.5}^{+0.3}$ degrees per $10^{5}$ yr for the `original', `stretched', and `new' shape variants respectively. Table 4 also summarises Apollo's spin change properties for the three different shape variants investigated.

\section{Discussion}

Considering that previous studies have shown the YORP effect to be highly sensitive to unresolved shape features and surface roughness (Statler 2009; Rozitis \& Green 2012), the shape model resolution (Breiter et al. 2009), and internal bulk density distribution (Scheeres \& Gaskell 2008); it is surprising that the theoretical YORP rotational acceleration predicted here matches the observed value very well. This is especially so when a light-curve-derived convex shape model was used for making the predictions. However, as pointed out in figure 5a of Rozitis \& Green (2013a), asteroids with rotation periods between 2 and 4 hours appear to have low degrees of concavity in their global shape. As Apollo has a rotation period of $\sim$3 hours, it suggests that it too may have a low degree of concavity, which is also supported by the low value of the convexity regularisation parameter that was used in the light-curve shape inversion. Rozitis \& Green (2013a) also showed that asteroids with high YORP-coefficients (i.e. >0.02) are less sensitive to the inclusion of concavities in their global shape model. From the derived physical properties and measured YORP rotational acceleration, Apollo has a high YORP-coefficient value of $\sim$0.03 (Rozitis \& Green 2013b), which suggests that its convex shape model could produce a reasonably accurate prediction. This level of insensitivity to shape model detail is also highlighted by the fact that when the roughness is allowed to vary in an extreme way across the surface (see Rozitis \& Green 2012), it only introduces an uncertainty of $\sim$10\% in the YORP effect predictions. Furthermore, if Apollo has a fractured rather than a rubble-pile interior (see below) then it is less likely for inhomogeneities in internal bulk density to occur.

Assuming a typical bulk density of $\sim$3330 kg m$^{-3}$ for the ordinary chondrites that are associated with S and Q-type asteroids (Carry 2012) allows the macro-porosity for Apollo to be estimated. The bulk density of 2400 $_{-470}^{+450}$ kg m$^{-3}$ derived for the `original' shape variant gives a macro-porosity of 28 $\pm$ 14 \%, which is on the border-line of Apollo having a rubble-pile interior (Britt et al. 2002). However, the higher bulk densities of 2820 $_{-660}^{+480}$ and 2850 $_{-680}^{+480}$ kg m$^{-3}$ derived for the `stretched' and `new' shape variants give macro-porosities of 15 $_{-14}^{+20}$ and 14 $_{-14}^{+21}$ \% respectively, which suggests a fractured interior like most S-type asteroids. A fractured interior suggests that Apollo could have enough coherent strength to prevent global-scale shape deformation caused by continued YORP spin-up. However, loose material resting on the surface of Apollo, as inferred from its low thermal inertia value, could be lost by continued YORP spin-up, and could explain the formation of Apollo's recently discovered Moon (Ostro et al. 2005).

The radar observations (Ostro et al. 2002) indicate Apollo's average OC radar cross section to be within 25\% of 0.24 km$^{2}$. Assuming the effective diameter derived for the `new' shape variant gives Apollo's radar albedo as 0.13 $\pm$ 0.03, a number which falls near the middle of values reported for other near-Earth asteroids (Benner 2012, NEA Radar Albedo Ranking\footnote{echo.jpl.nasa.gov/$\sim$lance/asteroid\_radar\_properties/nea.radaralbedo.html}). The radar circular polarisation ratio is sensitive to decimetre scale roughness, and Apollo's circular polarisation ratio has been measured to be 0.33 $\pm$ 0.01 indicating substantial surface roughness (Ostro et al. 2002). This is in agreement with the moderate degree of surface roughness inferred from the thermal-infrared observations since the radar wavelength scale of 12.6 cm is greater than the minimum surface roughness scale of $\sim$1 cm dictated by Apollo's implied diurnal thermal skin depth.

At their current rates, the YORP effect will halve Apollo's rotation period and shift its rotation axis to the asymptotic state at 180\degr~obliquity in just 2 to 3 Myr, whilst in the same period of time the Yarkovsky effect will decrease Apollo's semi-major axis by just $\sim$$10^{-3}$ AU. If most asteroids exhibit a YORP effect that acts on much shorter timescales than the Yarkovsky effect then it raises questions on the effectiveness of the Yarkovsky effect to transport main-belt asteroids into resonances leading to Earth crossing orbits. A main-belt asteroid must therefore perform a `semi-random walk' in orbital drift (e.g. Morbidelli \& Vokrouhlický 2003)  or reach a stable spin state, perhaps as a result of YORP-induced shape deformation (e.g. Cotto-Figueroa et al. 2012), to reach such resonance regions. A unified model, as presented here, is required to investigate the effectiveness of this process.

\section{Summary and conclusions}

We have investigated whether Apollo’s observed orbital semi-major axis drift and rotational acceleration are caused by the Yarkovsky and YORP effects respectively. The ATPM, which includes the effects of thermal-infrared beaming caused by unresolved surface roughness, has been demonstrated to consistently reproduce both effects, the thermal-infrared spectrum, and other observations using a single set of model parameters. This makes Apollo the fourth asteroid with a model confirmed Yarkovsky effect and, more importantly, the first asteroid with model confirmed detections of both the Yarkovsky and YORP effects.

In particular, ATPM derived an effective diameter of 1.55 $\pm$ 0.07 km, a geometric albedo of 0.20 $\pm$ 0.02, a thermal inertia value of 140 $_{-100}^{+140}$ J m$^{-2}$ K$^{-1}$ s$^{-1/2}$, and a roughness fraction of 60 $\pm$ 30 \%. This level of thermal inertia suggests that Apollo has loose regolith material resting on its surface, which is consistent with the view that Apollo has undergone recent resurfacing based on its observed Q-type spectrum. This degree of surface roughness is also consistent with that inferred from circular polarisation radar observations, which suggest that substantial centimetre-decimetre scale roughness exists on the surface. Furthermore, the `original' light-curve-derived shape model of Apollo required stretching in its z-axis by a factor of 1.17 in order to reconcile the thermal-infrared and radar measurements of its maximum equatorial diameter, and a `new' shape model with axis ratios of 1.94:1.65:1.00 was derived.

Comparisons between Apollo's predicted and measured Yarkovsky semimajor axis drift derived a bulk density of 2850 $_{-680}^{+480}$ kg m$^{-3}$ and a macro-porosity of 14 $_{-14}^{+21}$ \%, which are values consistent with those determined for other S-type asteroids, and suggest that Apollo has a fractured interior. The YORP effect is predicted to be increasing Apollo's obliquity towards one of the YORP asymptotic states at 180\degr~at a rate of 1.5 $_{-0.5}^{+0.3}$ degrees per $10^{5}$ yr. In addition to the observed YORP rotational acceleration, this indicates that the YORP effect is acting on a much faster timescale than the corresponding Yarkovsky effect and will dominate Apollo's long term evolution.

Finally, the ATPM can readily be applied to other asteroids with similar observation data sets to characterise their thermophysical properties, and then determine their long term evolution from the Yarkovsky and YORP effects.\\

\begin{acknowledgements}
The authors acknowledge the financial support of the UK Science and Technology Facilities Council (STFC). We also like to thank Steve Chesley for stimulating discussions that improved the accuracy of ATPM in producing Yarkovsky and YORP effect predictions.
\end{acknowledgements}

\begin{appendix}
\section{The accuracy of numerically solving 1D heat conduction in Yarkovsky and YORP effect predictions}

When computing the temperature variation with time and depth for a given facet, the 1D heat conduction equation is typically numerically solved using a finite difference method (e.g. Rozitis \& Green 2011). Due to the discrete nature of the finite difference method a sufficient number of time and depth steps are required to resolve the temperature variations, and the maximum depth at which 1D heat conduction is evaluated must be sufficiently deep such that diurnal/seasonal temperature variations are negligible. For thermophysical models used in interpreting thermal-infrared observations of asteroids, the maximum depth at which 1D heat conduction is typically evaluated has temperature variations that are a factor of $\sim$10$^{-3}$ to 10$^{-6}$ smaller than those on the surface, and the entire depth is divided into $\sim$30 to 60 depth steps. Similarly, $\sim$360 to 720 time steps are typically used to sample one asteroid rotation/orbit, and this is usually sufficient for accurately determining thermophysical properties from thermal-infrared observations.

To investigate whether these assumptions still produced accurate Yarkovsky orbital drift and YORP obliquity shift predictions, we performed a small study by varying these solving parameters. It was found that the typical maximum depths and number of time steps used in solving 1D heat conduction were already sufficient to produce accurate predictions, as changes by a factor of 2 produced Yarkovsky orbital drift and YORP obliquity shift predictions that differed by <0.5\%. However, it was found that differences in the number of depth steps used in solving 1D heat conduction did introduce small prediction differences, which were up to $\sim$30\% in some cases. In particular, it was noticed that the Yarkovsky orbital drift and YORP obliquity shift predictions converged with increasing number of depth steps used (see Fig. 7). It is impractical to use a high number of depth steps to ensure accurate predictions, especially when surface roughness is considered, because an even higher number of time steps will be required to ensure stability of the finite difference method [see equation (30) of Rozitis \& Green (2011)]. To get around this problem, we evaluate the Yarkovsky orbital drift and YORP obliquity shift at three different numbers of depth steps (i.e. 40, 48, and 56 depth steps), $x_{40}$, $x_{48}$, and $x_{56}$, and estimate the convergence limit (see horizontal grey lines of Fig. 7), $x_{\infty}$, using Aitken's delta-squared process
\begin{equation}
x_{\infty}=\frac{x_{56}x_{40}-x_{48}^{2}}{x_{56}-2x_{48}+x_{40}}\text{ .}
\label{(12)}
\end{equation}
The Yarkovsky and YORP effect predictions presented in Section 3 all utilised Aitken's delta-squared process and so they are as accurate as possible. Furthermore, the degree at which the Yarkovsky orbital drift was enhanced, and also the degree at which the YORP obliquity shift was dampened, by rough surface thermal-infrared beaming was not affected by the number of depth steps used. The main conclusions of Rozitis \& Green (2012) therefore remain unchanged.

\end{appendix}

\onecolumn

\begin{table}
\caption{Summary of (1862) Apollo radar (Ostro et al. 2002) and thermal-infrared (Lebofsky et al. 1981) observations obtained in 1980.}
\label{Table 1}
\centering
\begin{tabular}{l c c c c c c}
\hline\hline
Observation & Wavelength & Flux & Heliocentric & Geocentric & Phase angle & Sub-Earth \\
date & ($\muup$m) & ($10^{-14}$ & distance & distance & (\degr) & latitude (\degr) \\
(1980 UT) & & W m$^{-2}$ $\muup$m$^{-1}$) & (AU) & (AU) & & \\
\hline
Nov 14.92 & Radar & N/A & 0.996 & 0.057 & 81.8 & 23.4 \\
\\
Nov 17.26 & 10.2 & 7.62 $\pm$ 0.35 & & & & \\
Nov 17.27 & 20.0 & 3.53 $\pm$ 0.49 & 1.018 & 0.067 & 62.3 & 17.0 \\
\\
Nov 26.32 & 10.2 & 2.94 $\pm$ 0.27 & & & & \\
Nov 26.32 & 8.7 & 4.42 $\pm$ 0.61 & & & & \\
Nov 26.33 & 12.5 & 3.04 $\pm$ 0.42 & 1.105 & 0.148 & 34.9 & 2.8 \\
Nov 26.33 & 20.0 & 1.27 $\pm$ 0.12 & & & & \\
Nov 26.34 & 4.8 & 1.41 $\pm$ 0.20 & & & & \\
\\
Dec 4.14 & 10.6 & 1.24 $\pm$ 0.11 & 1.178 & 0.233 & 30.9 & -1.3 \\
\hline
\end{tabular}
\end{table}

\begin{table}
\caption{Assumed thermophysical modelling parameters for thermal-infrared flux fitting and Yarkovsky and YORP effect modelling.}
\label{Table 2}
\centering
\begin{tabular}{l c}
\hline\hline
Property & Value \\
\hline
Number of vertices & 1022 \\
Number of facets & 2040 \\
Rotation period\tablefootmark{a} & 3.065448 hrs \\
Pole orientation\tablefootmark{a} & $\lambda=48\degr, \beta=-72\degr$ \\
Obliquity\tablefootmark{a} & 162\degr \\
Semimajor axis\tablefootmark{b} & 1.47 AU \\
Eccentricity\tablefootmark{b} & 0.56 \\
Absolute magnitude\tablefootmark{c} & 16.384 \\
Phase parameter\tablefootmark{c} & 0.24 \\
Emissivity & 0.9 \\
\hline
\end{tabular}
\tablefoot{Obtained from \tablefoottext{a}{Ďurech et al. (2008a)}, the \tablefoottext{b}{JPL Small-Body Database Browser}, and \tablefoottext{c}{Pravec et al. (2012)}.}
\end{table}

\begin{table}
\caption{ATPM derived properties of (1862) Apollo using the Lebofsky et al. (1981) thermal-infrared observations at the 3-$\sigma$ confidence level.}
\label{Table 3}
\centering
\begin{tabular}{l c c c}
\hline\hline
Shape variant & `Original' & `Stretched' & `New' \\
\hline
Axis ratios ($a$:$b$:$c$) & 2.16:1.82:1.00 & 1.83:1.54:1.00 & 1.94:1.65:1.00 \\
$A/D$ ratio & 1.45 & 1.37 & 1.37 \\
Light-curve relative-$\chi^{2}$ & 2.59 & 3.32 & 2.35 \\
Effective diameter\tablefootmark{a} (km) & 1.68 $\pm$ 0.09 & 1.54 $\pm$ 0.07 & 1.55 $\pm$ 0.07 \\
Radar observed maximum & & & \\
equatorial diameter\tablefootmark{b} (km) & 2.07 $\pm$ 0.07 & 2.07 $\pm$ 0.07 & 2.07 $\pm$ 0.07 \\
ATPM maximum & & & \\
equatorial diameter\tablefootmark{a} (km) & 2.43 $\pm$ 0.13 & 2.12 $\pm$ 0.09 & 2.13 $\pm$ 0.09 \\
Geometric albedo\tablefootmark{a} & 0.17 $\pm$ 0.02 & 0.21 $\pm$ 0.02 & 0.20 $\pm$ 0.02 \\
Thermal inertia\tablefootmark{c} (J m$^{-2}$ K$^{-1}$ s$^{-1/2}$) & 200 $_{-140}^{+220}$ & 140 $_{-100}^{+160}$ & 140 $_{-100}^{+140}$ \\
Roughness fraction\tablefootmark{a} (\%) & 60 $\pm$ 30 & 60 $\pm$ 30 & 60 $\pm$ 30 \\
\hline
\end{tabular}
\tablefoot{\tablefoottext{a}{Mean and standard deviation.} \tablefoottext{b}{Measurement and uncertainty.} \tablefoottext{c}{Median and 1-$\sigma$ spread.}}
\end{table}

\begin{table}
\caption{Mass and spin change properties of (1862) Apollo derived by ATPM at the 3-$\sigma$ confidence level.}
\label{Table 4}
\centering
\begin{tabular}{l c c c}
\hline\hline
Shape variant & `Original' & `Stretched' & `New' \\
\hline
Measured Yarkovsky & & & \\
semimajor axis drift\tablefootmark{a} (m yr$^{-1}$) & -32.1 $\pm$ 3.4 & -32.1 $\pm$ 3.4 & -32.1 $\pm$ 3.4 \\
Bulk density\tablefootmark{b} (kg m$^{-3}$) & 2400 $_{-470}^{+450}$ & 2820 $_{-660}^{+480}$ & 2850 $_{-680}^{+480}$ \\
Macro-porosity\tablefootmark{b} (\%) & 28 $\pm$ 14 & 15 $_{-14}^{+20}$ & 14 $_{-14}^{+21}$ \\
Mass\tablefootmark{b} ($10^{12}$ kg) & 6.06 $_{-1.46}^{+1.05}$ & 5.49 $_{-1.58}^{+1.06}$ & 5.68 $_{-1.72}^{+1.14}$ \\
Moment of inertia\tablefootmark{b} ($10^{18}$ kg m$^{2}$) & 2.91 $_{-0.95}^{+0.70}$ & 1.98 $_{-0.68}^{+0.52}$ & 2.17 $_{-0.77}^{+0.60}$ \\
Measured YORP & & & \\
rotational acceleration\tablefootmark{c} ($10^{-3}$ rad yr$^{-2}$) & 7.3 $\pm$ 1.6 & 7.3 $\pm$ 1.6 & 7.3 $\pm$ 1.6 \\
Predicted YORP & & & \\
rotational acceleration\tablefootmark{b} ($10^{-3}$ rad yr$^{-2}$) & 6.8 $_{-1.2}^{+2.7}$ & 6.8 $_{-1.3}^{+3.5}$ & 6.1 $_{-1.2}^{+3.2}$ \\
YORP obliquity shift\tablefootmark{b} (\degr~/ $10^{5}$ yr) & 2.4 $_{-0.6}^{+0.5}$ & 1.7 $_{-0.5}^{+0.3}$ & 1.5 $_{-0.5}^{+0.3}$ \\
\hline
\end{tabular}
\tablefoot{\tablefoottext{a}{Mean and standard error.} \tablefoottext{b}{Median and 1-$\sigma$ spread.} \tablefoottext{c}{Measurement and uncertainty.}}
\end{table}

\begin{figure}
\centering
\includegraphics[width=\hsize/2]{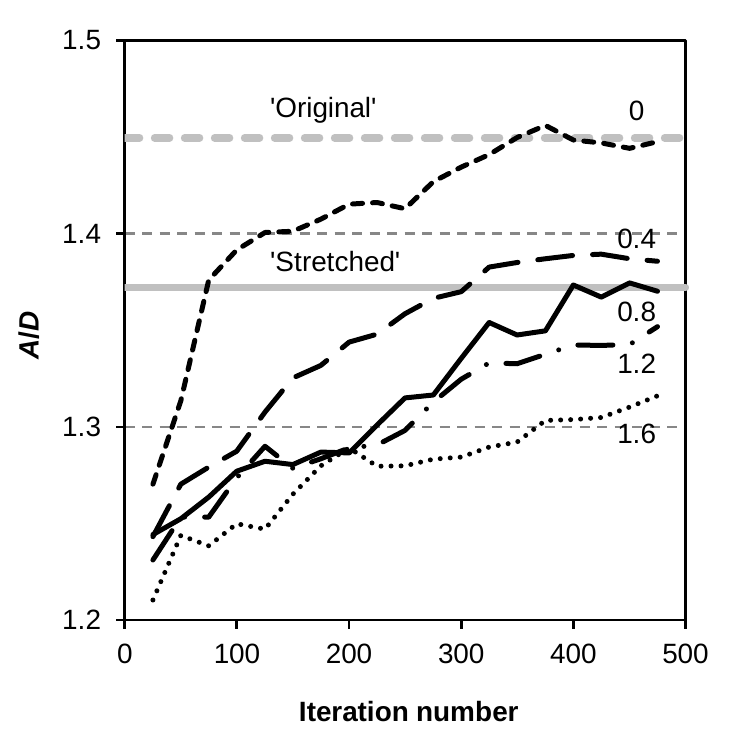}
\caption{Maximum equatorial diameter to effective diameter ratio ($A/D$) of the (1862) Apollo light-curve-derived shape models as a function of iteration number (x-axis) and convexity regularisation parameter (different line styles). The dashed and solid horizontal grey lines correspond to the $A/D$ values determined for the `original' and `stretched' shape model variants, respectively, for comparison purposes.}
\label{Fig. 1.}
\end{figure}

\begin{figure}
\centering
\includegraphics[width=\hsize]{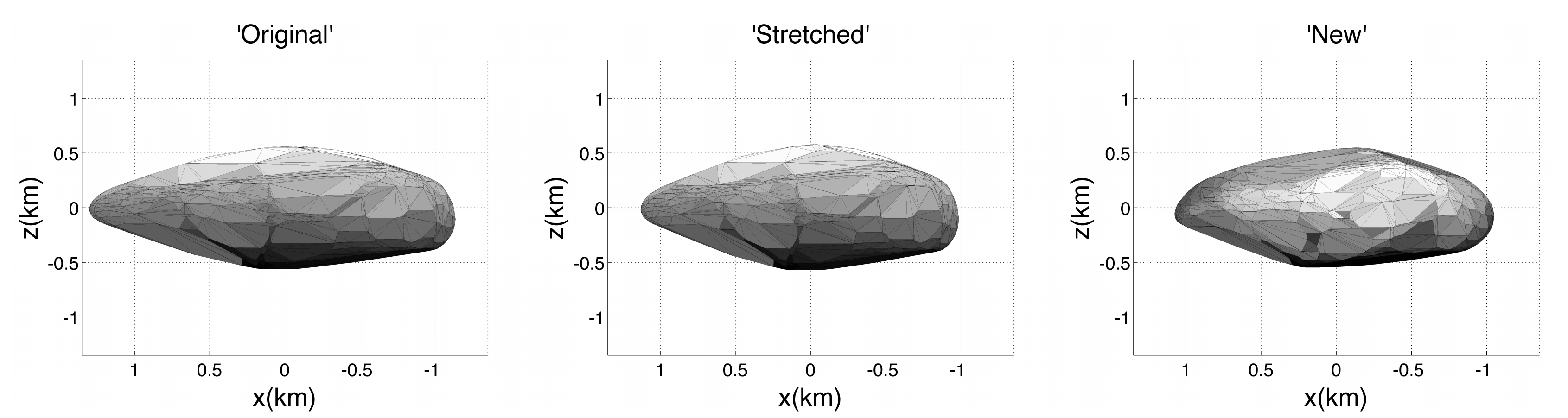}
\caption{`Original', `stretched', and `new' light-curve-derived shape model variants of (1862) Apollo respectively. They have been scaled to their mean sizes determined by the ATPM thermal-infrared flux fitting.}
\label{Fig. 2.}
\end{figure}

\begin{figure}
\centering
\includegraphics[width=\hsize/2]{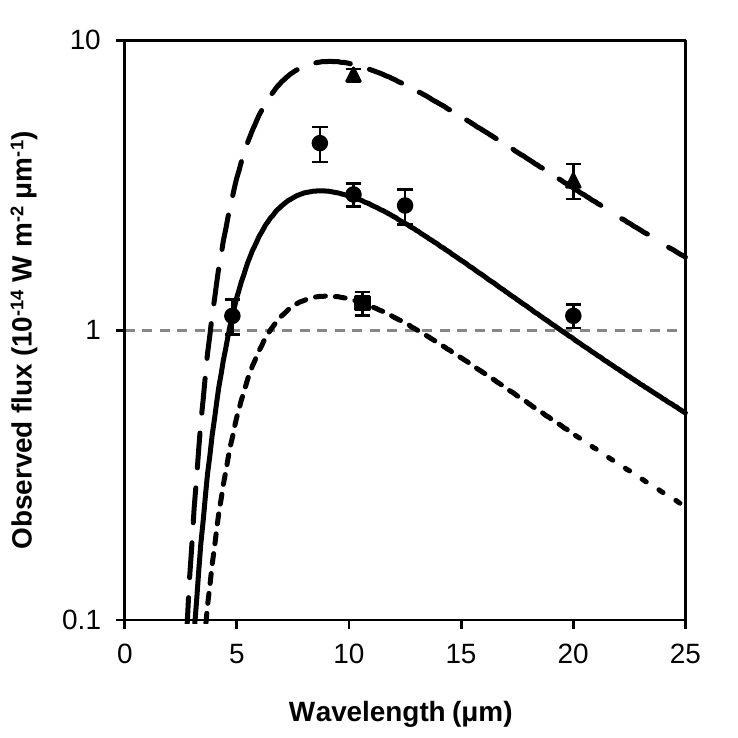}
\caption{Example ATPM fit to the (1862) Apollo thermal-infrared observations using the `new' shape model variant, a thermal inertia of 140 J m$^{-2}$ K$^{-1}$ s$^{-1/2}$, and a roughness fraction of 60\%. Thermal-infrared observations obtained on the 17th November, 26th November, and 4th December 1980 are depicted by the solid triangles, circles, and square (Lebofsky et al. 1981), and their model fits are represented by the dashed, solid, and dotted lines respectively. The model thermal-infrared spectra are shown at a single rotation phase corresponding to when the $\sim$10 $\muup$m observation was conducted for each observation set. However, the observations were made at different rotation phases within each observation set and therefore have been adjusted to a single rotation phase for direct visual comparison with the model. To adjust the observations, the ratios of the model fluxes from their respective rotation phases to the $\sim$10 $\muup$m observation rotation phase was calculated and then applied to each flux measurement. This then allows direct visual comparison of the model thermal-infrared spectrum and the observations at a single rotation phase. However, for ATPM chi-square fitting, the observed fluxes were compared to the model fluxes at their respective observation rotation phases.}
\label{Fig. 3.}
\end{figure}

\begin{figure}
\centering
\includegraphics[width=\hsize]{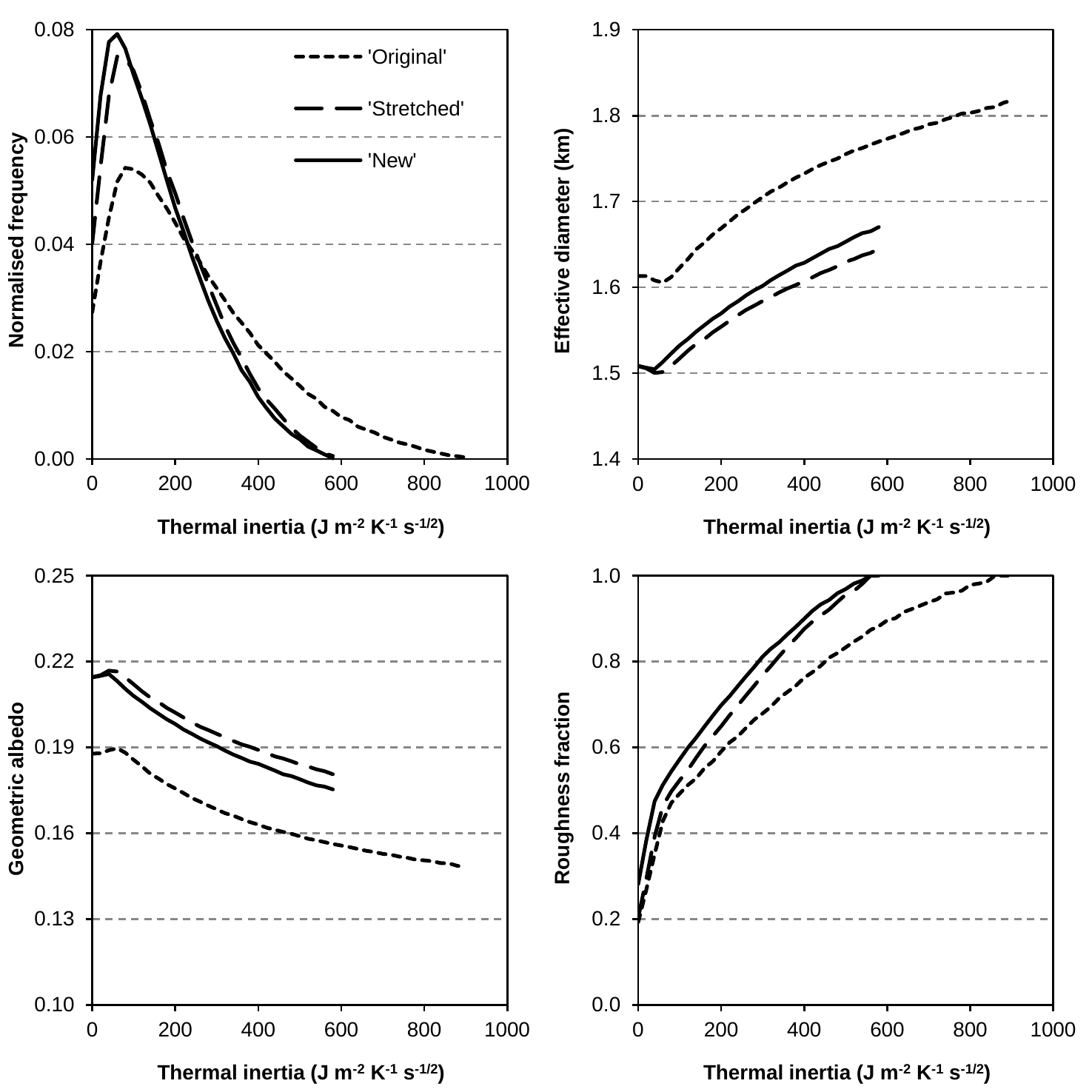}
\caption{Summary of ATPM chi-square fitting results to the Lebofsky et al. (1981) thermal-infrared observations of (1862) Apollo. The possible thermal inertia distribution (top left), and the co-variance with thermal inertia of the average effective diameter (top right), geometric albedo (bottom left), and roughness fraction (bottom right) derived at the 3-$\sigma$ confidence level for each of the three different shape model variants (legend).}
\label{Fig. 4.}
\end{figure}

\begin{figure}
\centering
\includegraphics[width=\hsize]{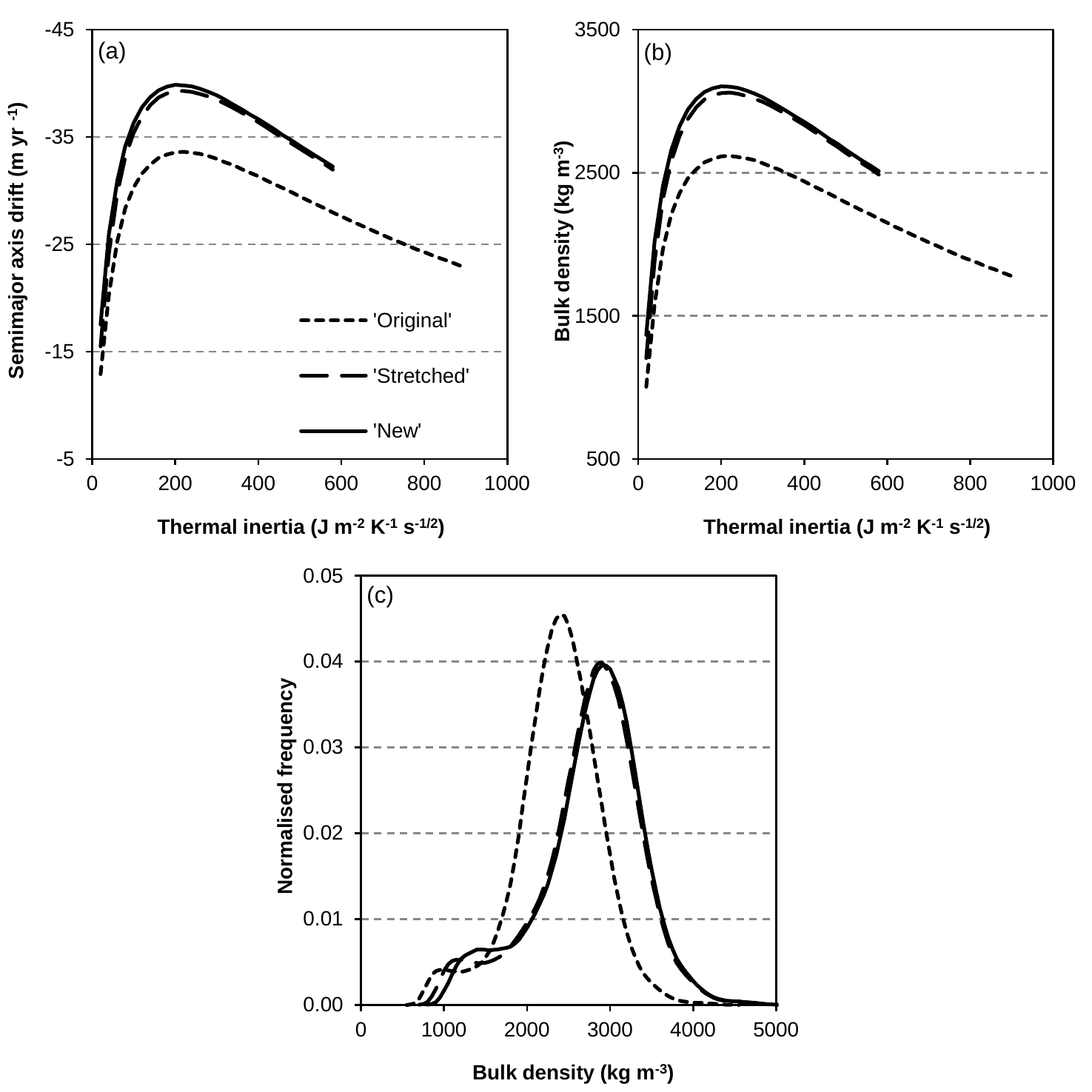}
\caption{Summary of ATPM Yarkovsky effect modelling results for (1862) Apollo using its three different shape model variants (legend). (a) Average Yarkovsky semimajor axis drift as a function of thermal inertia at a fixed bulk density of 2500 kg m$^{-3}$. (b) Average bulk density as a function of thermal inertia derived by comparing the model orbital drift against that measured. (c) The distribution of possible bulk densities derived at the 3-$\sigma$ confidence level.}
\label{Fig. 5.}
\end{figure}

\begin{figure}
\centering
\includegraphics[width=\hsize]{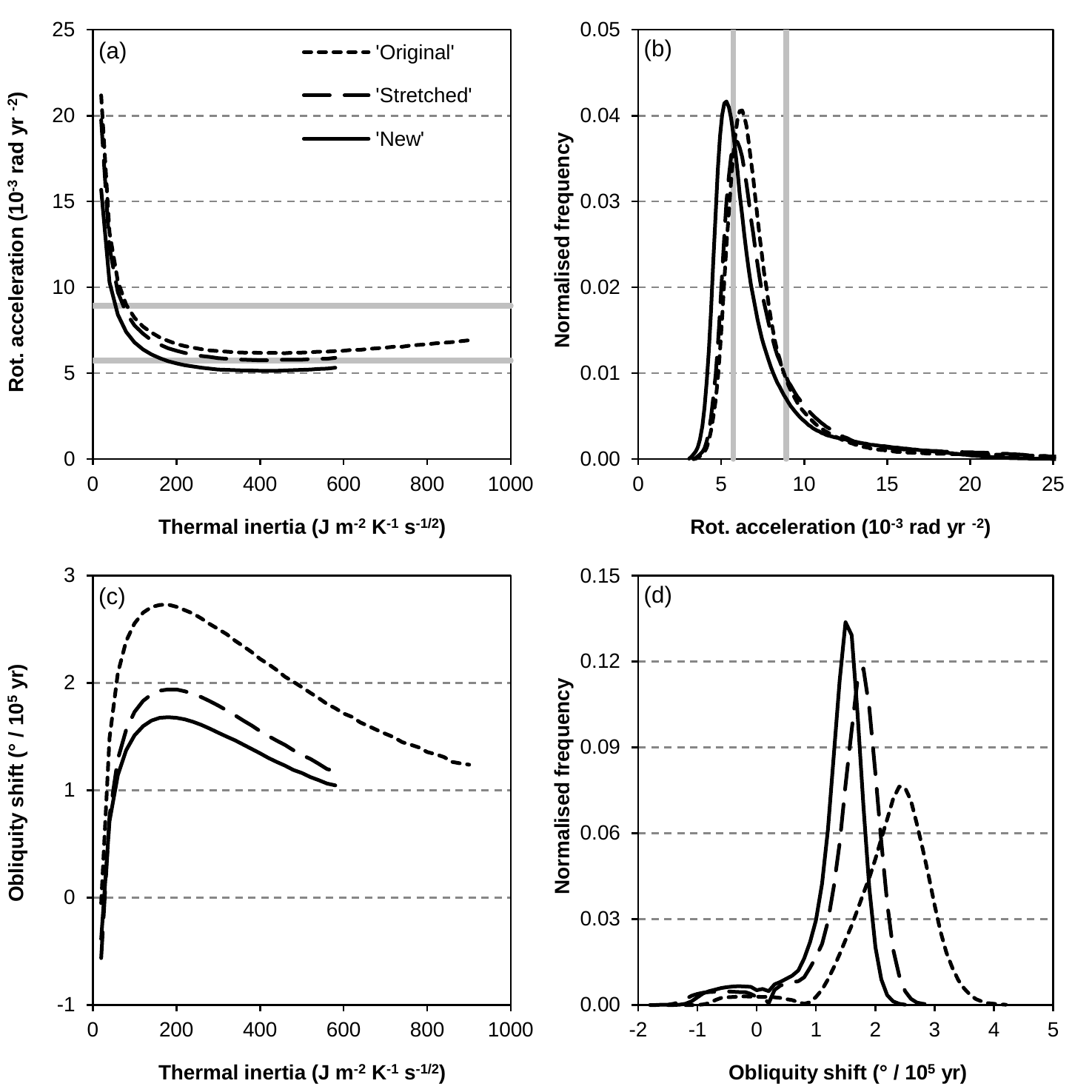}
\caption{Summary of ATPM YORP effect modelling results for (1862) Apollo using its three different shape model variants (legend). (a) Average YORP rotational acceleration as a function of thermal inertia. (b) The distribution of possible YORP rotational acceleration values derived at the 3-$\sigma$ confidence level. (c) Average rate of YORP obliquity shift as a function of thermal inertia. (d) The distribution of possible rates of YORP obliquity shift derived at the 3-$\sigma$ confidence level. The grey lines shown in panels (a) and (b) represent the lower and upper bounds of the YORP rotational acceleration acting on (1862) Apollo as measured by Ďurech et al. (2008a).}
\label{Fig. 6.}
\end{figure}

\begin{figure}
\centering
\includegraphics[width=\hsize/2]{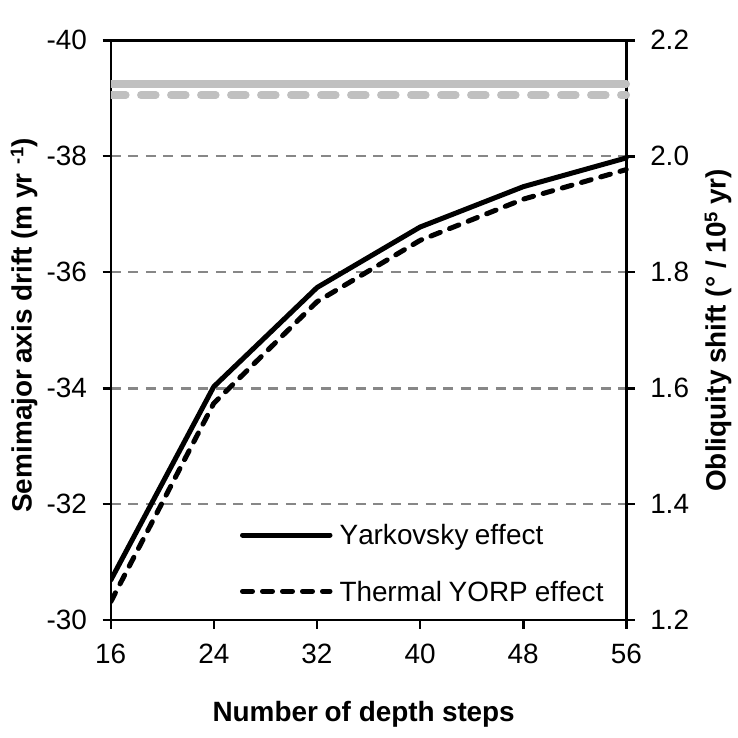}
\caption{Convergence of the predicted Yarkovsky semimajor axis drift (solid line plotted on primary y-axis) and thermal component of the YORP obliquity shift (dashed line plotted on secondary y-axis) with number of depth steps used to solve the 1D heat conduction equation. The horizontal grey lines represent the convergence limits estimated by Aitken's delta-squared process. The example convergence series shown here are calculated for the `new' shape model variant of (1862) Apollo assuming a thermal inertia of 140 J m$^{-2}$ K$^{-1}$ s$^{-1/2}$ and a roughness fraction of 60\%.}
\label{Fig. A1.}
\end{figure}

\end{document}